\documentclass[twocolumn,showpacs,preprintnumbers,amsmath,amssymb,superscriptaddress]{revtex4}

\usepackage{epsfig}
\usepackage{graphicx}
\usepackage{dcolumn}
\usepackage{bm}

\newcommand{\be}{\begin{equation}}
\newcommand{\ee}{\end{equation}}

\begin{document}

\preprint{APS/123-QED}

\title{Atom interferometers with scalable enclosed area}
\author{Holger M\"uller}
\affiliation{Department of Physics, 366 Le Conte Hall, University
of California, Berkeley, CA 94720-7300} \affiliation{Lawrence
Berkeley National Laboratory, one Cyclotron Road, Berkeley, CA
94720.} \altaffiliation{Physics Department, Stanford University,
382 Via Pueblo Mall, Stanford, California 94305, USA}
\email{hm@berkeley.edu}
\author{Sheng-wey Chiow}
\author{Sven Herrmann}
\affiliation{Physics Department, Stanford University, 382 Via
Pueblo Mall, Stanford, California 94305, USA}
\author{Steven Chu}
\affiliation{Department of Physics, 366 Le Conte Hall, University
of California, Berkeley, CA 94720-7300} \affiliation{Lawrence
Berkeley National Laboratory, one Cyclotron Road, Berkeley, CA
94720.} \altaffiliation{Physics Department, Stanford University,
382 Via Pueblo Mall, Stanford, California 94305, USA}

\date{\today}

\begin{abstract}
Bloch oscillations (i.e., coherent acceleration of matter waves by
an optical lattice) and Bragg diffraction are integrated into
light-pulse atom interferometers with large momentum splitting
between the interferometer arms, and hence enhanced sensitivity.
Simultaneous acceleration of both arms in the same internal states
suppresses systematic effects, and simultaneously running a pair
of interferometers suppresses the effect of vibrations.
Ramsey-Bord\'e interferometers using four such Bloch-Bragg-Bloch
(BBB) beam splitters exhibit 15\% contrast at 24$\hbar k$
splitting, the largest so far ($\hbar k$ is the photon momentum);
single beam splitters achieve 88$\hbar k$. The prospects for
reaching 100s of $\hbar k$ and applications like gravitational
wave sensors are discussed.
\end{abstract}

 \maketitle

\begin{figure*}\centering
\epsfig{file=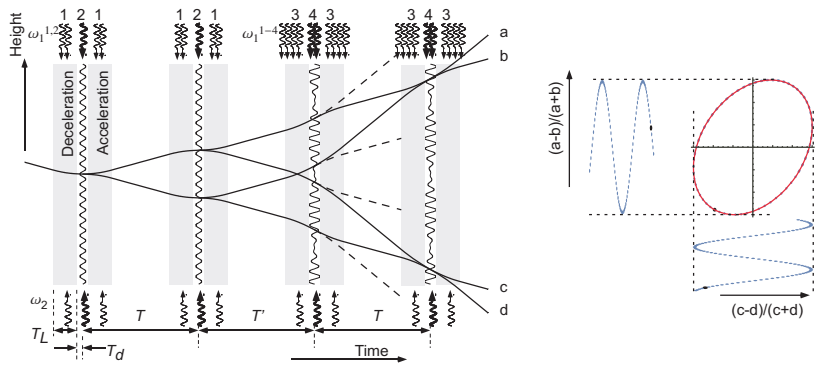,width=0.6\textwidth}
\caption{\label{interferometer} Left: Space-Time diagram of
simultaneous conjugate Ramsey-Bord\'e BBB-Interferometers. 1: Dual
optical lattice; 2: Single Bragg beam splitter; 3: Quadruple
optical lattice; 4: Dual Bragg beam splitter; a-d: outputs. The
dashed lines indicate trajectories that do not interfere. Right:
plotting the normalized populations of the outputs $(a-b)/(a+b)$
versus $(c-d)/(c+b)$ draws an ellipse.}
\end{figure*}

Light-pulse atom interferometers use the momentum transfer in
light-atom interactions to split and reflect matter waves, thus
forming the interferometer arms \cite{Pritchardreview}. They have
been used for measurements of high precision and accuracy, such as
of gravity \cite{Peters}, the fine-structure constant
\cite{Weiss,Wicht,Biraben}, gravity gradients \cite{Snaden98},
Newton's gravitational constant \cite{Fixler,Lamporesi}, and one
of the few terrestrial tests of general relativity that is
competitive with astrophysics \cite{LVGrav}. In these examples,
the atom-light interactions are Raman transitions which transfer
the momentum $\Delta p=2\hbar k$ of two photons. This limits the
space-time area enclosed between the interferometer arms, and
hence the sensitivity. Bragg diffraction of matter waves has been
used to increase $\Delta p$ \cite{Bragg}, the current record being
$24\hbar k$ \cite{BraggInterferometry,SCI}. Unfortunately, this
number represents a technical limit, as the required laser power
increases sharply with $\Delta p$ \cite{Losses}. Bloch
oscillations \cite{Peik,Biraben,Ferrari} or adiabatic transfer
\cite{Weiss,Wicht} have been used to transfer thousands of $\hbar
k$, but so far only to the common momentum of the interferometer
arms, which does not increase the enclosed area. Here, we use them
to increase the splitting of the arms and thus increase the
enclosed  area (Fig. \ref{interferometer}). Bragg beam splitters
are embedded between Bloch oscillations to achieve a $\Delta p$ of
up to $88\hbar k$. In Ramsey Bord\'e interferometers with $\Delta
p=24\hbar k$ , we see interferences with $\sim 15\%$ contrast
(compared to 4\% with Bragg diffraction
\cite{BraggInterferometry}). The enclosed area is no longer
limited by available laser power. This work is thus the first
demonstration of an interferometer whose enclosed area can be
scaled up to allow for proposed landmark experiments such as
detection of low-frequency gravitational waves \cite{GravWav} or
the Lense-Thirring effect \cite{Landragin}, tests of the
equivalence principle at sensitivities of up to $\delta g/g\sim
10^{-17}$ \cite{Dimopoulos}, atom neutrality \cite{Neutr}, or
measurements of fundamental constants with sensitivity to
supersymmetry \cite{Paris}.

The two basic ingredients of this work are Bragg diffraction of
matter waves at an optical lattice and Bloch oscillations of
matter waves in an accelerated optical lattice. To describe them,
we consider an atom of mass $M$ in the electric fields of two
counterpropagating laser beams whose frequencies are $\omega_1$
and $\omega_2$. We denote $k=(\omega_1+\omega_2)/(2c)$ the average
wavenumber.

For Bragg diffraction \cite{Losses}, $\omega_1$ and $\omega_2$ are
constant in the rest frame of the atom. Neglecting spontaneous
processes, the atom absorbs $n$ photons at $\omega_1$ from one
beam and is stimulated to emit $n$ into the other beam at
$\omega_2$. The atom emerges in its original internal quantum
state but moving with a momentum of $2n\hbar k$ and a kinetic
energy of $(2n\hbar k)^2/(2M)\equiv 4n^2\hbar \omega_r$, where
$\omega_r$ is the recoil frequency. This must match the energy
$n\hbar(\omega_1-\omega_2)$ lost by the laser field, which
determines the Bragg diffraction order $n$. To make a beam
splitter for matter waves, the pulse duration and intensity are
chosen such that the process happens with a probability of 1/2 (a
``$\pi/2$"-pulse).


For Bloch oscillations, suppose the beams initially have zero
difference frequency, which is then ramped linearly with time,
$\omega_1-\omega_2 = \dot\omega t$. At a certain time, it will
thus satisfy the Bragg resonance condition for transitions between
momentum states $\left|p=0\right> \rightarrow \left|p=2\hbar
k\right>$ and then subsequently $\left|p=2\hbar k\right>
\rightarrow \left|p=4\hbar k\right>, \ldots $. The atom thus
receives $2\hbar k$ of momentum in intervals given by the Bloch
period $\tau_B= 8\omega_r/\dot\omega $. Theory and experiment show
that when $\dot \omega$ satisfies an adiabaticity criterion
\cite{Peik} and the difference frequency is held constant after it
reaches $\omega_1-\omega_2=8N\omega_r$, where $N=1,2,3,\ldots $,
the population emerges in an $\left|p=2N\hbar k \right>$ momentum
state which is nearly pure \cite{Peik}.

The simplest way to increase the momentum transfer of a beam
splitter with Bloch oscillations would be to take one output of a
conventional beam splitter and accelerate it with Bloch
oscillations \cite{Hecker}; a total momentum splitting of $10\hbar
k$ has been achieved \cite{CladeBOInterferometer} with a contrast
at the few-percent level. According to simulations, parameters
like the lattice depth and acceleration can be chosen such that
the other arm can remain un-accelerated if the initial momentum
splitting is sufficient \cite{HauganKovachy}. In this
configuration, however, the lattice causes an imbalance in the ac
Stark effect between arms. This can easily contribute tens of
radians to the interferometer phase as a function of the lattice
depth. Even if this can be cancelled between subsequent optical
lattices \cite{CladeBOInterferometer}, fluctuations or drift in
the laser power, and hence the lattice depth, may spoil the mrad
precision \cite{Peters} of typical interferometers.

To minimize this systematic effect, it is better to simultaneously
accelerate both arms with a pair of lattices. This will balance
the ac Stark effect. The balance can be maintained by controlling
the intensity balance of the lattice beams, which is technically
much easier than accurate control of their absolute intensity. The
final version of our beam splitter also has a pair of Bloch
oscillations at the input, which decelerate the relative velocity
of the arms. This leads to symmetry with respect to interchanging
inputs and outputs, facilitating use in atom interferometers.

\begin{figure}\centering
\epsfig{file=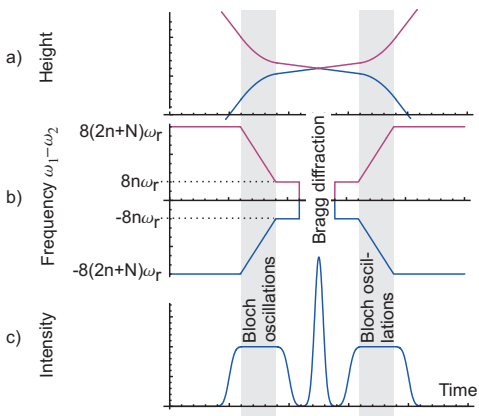,width=0.35\textwidth}
\caption{\label{BSschem} Trajectories of the atom (a), laser
frequencies (b), and laser intensities (c) during a BBB beam
splitter.}
\end{figure}

Fig. \ref{BSschem} shows our final configuration, which we may
call the Bloch-Bragg-Bloch (BBB) beam splitter. Consider the first
and second beam splitter of Fig. \ref{interferometer}. Initially,
its two atomic inputs have a momentum difference of $(2n+4N)\hbar
k$. Each is loaded into a co-moving optical lattice. The lattices
decelerate the momentum difference to $2n\hbar k$. Afterwards,
when the paths cross, a Bragg diffraction acts as a beam splitter.
Finally, the outputs are accelerated symmetrically to a splitting
of $(2n+4N)\hbar k$. To make the simultaneous accelerated
lattices, one can, for example, superimpose two frequencies
$\omega_1^{1,2}$ in one beam which are ramped as shown in Fig.
\ref{BSschem} b. The difference frequencies in this example start
at $(\omega_1^1-\omega_2)=-(\omega_1^2-\omega_2)=8(2N+n)\omega_r$
and are ramped down to $8n\omega_r$. For Bragg diffraction, there
is just one frequency in each beam, so that $\omega_1=\omega_2$ in
this specific reference frame. The figure also shows the laser
intensity versus time (the intensities of the counterpropagating
beams are proportional to each other) \cite{beatfreq}.

The realization of these ideas faces several challenges: For one,
the simultaneous use of two optical lattices to accelerate
populations coherently into different directions has never been
demonstrated. Moreover, each BBB beam splitter uses four optical
lattices and one Bragg diffraction, and a full interferometer
consists of four beam splitters. Even if single accelerated
lattices and Bragg diffractions have been shown to be coherent,
the end-to-end coherence of a BBB beam splitter or a full
interferometer may be very hard to realize. Moreover,
interferometers with very high momentum transfer are prone to loss
of interference contrast due to vibrational noise, even with
state-of-the-art vibration isolation \cite{BraggInterferometry}.

To avoid this loss of contrast, we use simultaneous conjugate
interferometers (SCI) \cite{SCI}. Two interferometers are run
simultaneously, and their fringes are plotted against each other,
forming an ellipse (Fig. \ref{interferometer}). Common-mode phase
fluctuations move the data around the ellipse, but do not affect
the eccentricity, which is determined by the differential-mode
phase. Ellipse-specific fitting \cite{ellipfit} or Bayesian
estimation \cite{Stockton} allows us to extract the differential
phase independent of common-mode noise. Whereas the first and
second beam splitter are common to our SCIs, the third and fourth
are two superimposed BBB splitters, each addressing one
interferometer according to the momentum of the atom (Fig.
\ref{interferometer}). Thus, the pair of SCIs even requires
quadruple optical lattices, for a total of 24 optical lattices and
6 Bragg diffractions. Coherence of such an intricate
atom-optics-system has rarely been demonstrated, if at all.


Our experiment uses a 1-m high atomic fountain of about $10^6$ Cs
atoms in the $F=3, m_F=0$ quantum state with a velocity
distribution of 0.3 recoil velocities full width at half maximum
(FWHM). It is based on a 3 dimensional magneto-optical trap (MOT)
with a moving optical molasses launch and subsequent Raman
sideband cooling in a co-moving optical lattice \cite{Treutlein}.

\begin{figure}\centering
\epsfig{file=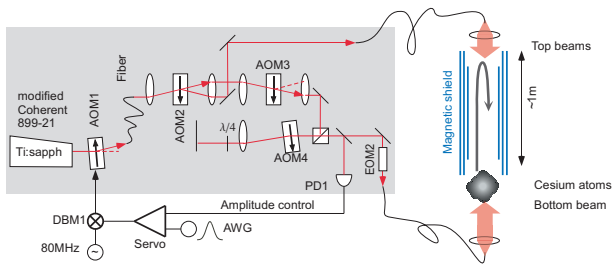,width=.45\textwidth}
\caption{\label{Setup} Setup. AWG: arbitrary waveform generator.
$\lambda/4$: quarter-wave retardation plate. DBM: double-balanced
mixer.}
\end{figure}

Our laser system for driving Bloch oscillations and Bragg
diffraction is based on a 6\,W Ti:sapphire laser
\cite{6Wlaser,SCI}, stabilized with a red detuning of 16\,GHz to
the $F=3\rightarrow F'=4$ transition in the cesium D2 manifold
near 852\,nm. An acousto-optical modulator AOM 1 (Fig.
\ref{Setup}) is used for closed-loop amplitude control. To
generate $\omega_1^{1-4}$, AOM 2 is driven with up to 4 radio
frequencies near 180\,MHz (these frequencies are close enough so
that the difference in deflection angle can be neglected). Phase
shifts due to optical path length fluctuations are thus common to
$\omega_1^{1-4}$ to a high degree. This is essential for the
cancellation of vibrations between the conjugate interferometers.
The beam at $\omega_2$ is ramped using the double-passed AOM 4.
Such ramping accounts for the free fall of the atoms, which
changes the resonance condition in the laboratory frame at a rate
of $\sim 23$\,MHz/s. The two beams are brought to the experiment
via two single-mode, polarization-maintaining fibers and
collimated to an $1/e^2$ intensity radius of about 3\,mm. The
bottom beam has a maximum power of 1.15\,W at the fiber output;
the top beam a peak power of 1.6\,W, or respectively 0.4\,W and
0.1\,W per frequency in two and four-frequency operation. See Ref.
\cite{SCI} for more details.

An experimental sequence starts with launching the atoms to a 1-s
vertical ballistic trajectory. The first 70\,ms are used for
preparing the atomic sample. After the experiment, the atoms are
fluorescence detected as they pass a photomultiplier tube (PMT).

\begin{figure}
\centering \epsfig{file=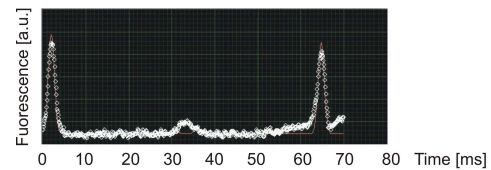,width=0.35\textwidth}
\caption{\label{88hbark} Time-of-flight sheet of the output of an
88$\hbar k$ beam splitter.}
\end{figure}

The Bragg pulses of our BBB beam splitters are Gaussian with a
$1/\sqrt{e}$ half-width $\sigma \sim 20\,\mu$s. Their intensity is
adjusted for a 50\% diffraction efficiency. The Bloch oscillation
acceleration phase has a duration of $1-2$\,ms. The laser
intensity is set sufficiently high for a Bloch oscillation
efficiency of nearly 100\% \cite{Peik} and is adiabatically
switched on and off with a rise time of several 100$\,\mu$s. Fig.
\ref{88hbark} shows the PMT fluorescence signal of the population
after one beam splitter versus time. Since the atoms reach the PMT
at a time dependent on their velocity after the beam splitter, the
x-axis gives the momentum transfer with a scaling of
0.7\,ms/$(\hbar k)$. The resolution of about $1\hbar k$ is
determined by the spatial extent of the atomic sample, and the
slots used for the detection beam and PMT. Fig. \ref{88hbark}
shows a momentum splitting of 88$\hbar k$, of which $8\hbar k$
have been transferred by Bragg diffraction. The small peak in the
middle is due to atoms which could not be diffracted.

\begin{figure}
\centering 
\epsfig{file=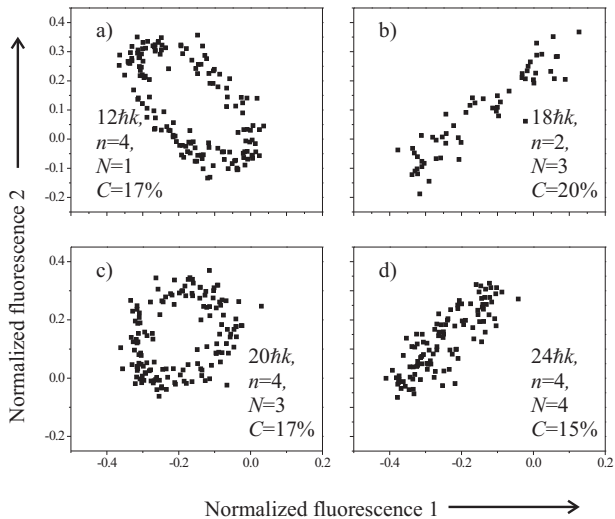,width=0.45\textwidth}
\caption{\label{simconj} Ellipses from simultaneous conjugate
interferometers with BBB beam splitters. Total momentum transfer,
Bragg diffraction order $n$, Bloch oscillation number $N$, and
contrast $C$ are stated. All graphs are plotted using the same
scale.}
\end{figure}

A pair of interferometers is driven by combining BBB splitters in
the way shown in Fig. \ref{interferometer}. The population in the
four outputs a-d is detected by their fluorescence $f_{a-d}$. To
take out fluctuations in the initial atom number, we define the
normalized fluorescence $F_u=(f_a-f_b)/(f_a+f_b)$ of the upper
interferometer and $F_l$ in analogy for the lower interferometer.
Fig. \ref{simconj} shows ellipses of the two interferometers,
obtained by plotting their interference fringes $F_u, F_l$ versus
one another. Interferometers with momentum transfers between
12-24$\hbar k$ are shown. The pulse separation time $T$ (see Fig.
\ref{interferometer}) was between 2-10\,ms, with little influence
on contrast as expected for simultaneous interferometers
\cite{SCI}. We usually transferred $(4-8)\hbar k$ by Bragg
diffraction, see Fig. \ref{simconj}. Even at the highest total
splitting of $24\hbar k$, a contrast of 15\% is achieved, compared
to 4\% in the only previous instrument with such splitting (based
on Bragg diffraction alone) \cite{BraggInterferometry}. This
contrast is 30\% of the theoretical value for Ramsey-Bord\'e
interferometers \cite{BraggInterferometry,SCI}.

The momentum splitting of $24\hbar k$ is currently limited by two
technical issues: (i) The frequency ramps for driving Bloch
oscillations were generated by a staircase approximation with at
most 1024 steps. Large spans thus require a coarse step size,
leading to loss of coherence. A better ramp generator will resolve
this issue. (ii) Distortions of the wavefronts of the laser beam,
caused by imperfect lenses, waveplates, and vacuum viewports, lead
to loss of contrast as described in \cite{SCI}. A mode-filtering
cavity will alleviate this.

Outside of interferometers, Bloch oscillations have already been
used to coherently transfer 8000$\hbar k$ \cite{Ferrari}. The BBB
splitter thus opens up the door to substantial increases in
atom-interferometer-sensitivity. One exciting application for such
an ultra-sensitive device is an atomic gravitational wave
interferometric sensor (AGIS) \cite{GravWav}. Such a sensor could
access the gravitational wave spectrum between 0.1-100\,Hz,
complementing the reach of optical interferometers such as LIGO
and LISA. Sensing in this spectral region enables searches for
white dwarf, black hole, and intermediate mass black hole
binaries, or for the stochastic gravitational wave background from
phase transitions in the early universe. A crucial ingredient of
an AGIS is a momentum splitting of $100\hbar k$ or more; another
is common-mode rejection of vibrations between simultaneous
interferometers. Both common-mode rejection as well as large
momentum transfer have already been demonstrated here. Other
applications requiring extremely large enclosed areas include
measuring the Lense-Thirring effect \cite{Landragin}, or tests of
the equivalence principle \cite{Dimopoulos} and atom neutrality
\cite{Neutr}.

To summarize, we used Bloch oscillations and Bragg diffraction to
demonstrate atom interferometers with a scalable momentum-space
splitting $\Delta p$ between the arms. Simultaneous Bloch
oscillations and use of a single internal state
\cite{BraggInterferometry} in both arms suppress systematic
effects; simultaneous interferometers \cite{SCI} suppress
vibrational noise. Individual beam splitters reach $\Delta
p=88\hbar k$; interferometers with $\delta p=24\hbar k$ reach 30\%
of their optimum contrast. This represents the highest momentum
transfer realized to date in any light pule atom interferometer
\cite{BraggInterferometry}, while substantially improving
contrast. Most importantly, $\delta p$ can be scaled up, as it is
no longer limited by the available laser power. This will be
instrumental in realizing several proposed landmark experiments
\cite{GravWav,Landragin,Dimopoulos,Neutr,Paris}. For example, ours
is the first experimental demonstration of atom interferometers
that can be scaled up for gravitational wave detection.

We are indebted to Jason Hogan, Mark Kasevich, Tim Kovachy, and
Shau-yu Lan for discussions. This material is based upon work
supported by the National Science Foundation under Grant No.
0400866. S.H. and H.M. thank the Alexander von Humboldt
Foundation.


\begin{references}

\bibitem{Pritchardreview} A. D. Cronin, J. Schmiedmayer, and D. E.
Pritchard, e-print arXiv:0712.3703v1 (submitted to Rev. Mod.
Phys.)

\bibitem{Peters} A. Peters, K. Y. Chung, and S. Chu, Nature (London) {\bf 400,} 849 (1999).

\bibitem{Weiss} D. S. Weiss,
B. C. Young, and S. Chu, \prl {\bf 70,} 2706 (1993); M. Weitz, B.
C. Young, and S. Chu, {\em ibid.} {\bf 73,} 2563 (1994).

\bibitem{Wicht}
A. Wicht {\em et al.}, Physica Scripta {\bf T102,} 82 (2002).

\bibitem{Biraben} P. Clad\'{e} {\em et al.,} \prl {\bf 96,}
033001 (2006); Phys. Rev. A {\bf 74,} 052109 (2006); M. Cadoret
{\em et al.,} arXiv:0809.3177.

\bibitem{Snaden98} M. J. Snadden {\em et al.}, \prl {\bf 81,} 971 (1998).


\bibitem{Fixler} J. B. Fixler {\em et al,} Science {\bf 315,} 74 (2007).

\bibitem{Lamporesi} G. Lamporesi {\em et al.}, Phys. Rev. Lett. {\bf 100,} 050801 (2008).

\bibitem{LVGrav} H. M\"uller {\em et al.}, \prl {\bf 100,} 031101 (2008).

\bibitem{Bragg} D. M. Giltner, R. W. McGowan, and S. A. Lee, \prl {\bf 75,} 2638
(1995); A. Miffre {\em et al.}, Eur. Phys. J. D {\bf 33,} 99
(2005).

\bibitem{BraggInterferometry} H. M\"uller {\em et al.}, \prl {\bf 100,} 180405 (2008).

\bibitem{SCI} S. Herrmann, S.-w. Chiow, S. Chu, and H. M\"uller, arXiv:0901.1819

\bibitem{Losses} H. M\"uller, S.-w. Chiow, and S. Chu, Phys. Rev. A {\bf 77,}
023609 (2008).

\bibitem{Peik} E. Peik {\em et al.}, Phys. Rev. A {\bf 55,} 2989
(1997).

\bibitem{Ferrari} G. Ferrari, {\em et al.},
\prl {\bf 97,} 060402 (2006).

\bibitem{GravWav} S. Dimopoulos {\em et al.}, Phys. Rev. D {\bf 78,} 122002
(2008).

\bibitem{Landragin} A. Landragin {\em et al.}, in: {\em Optics in Astrophysics,} edited by R. Foy and F.
C. Foy (NATO Science Series II, Vol. 198; Springer, Berlin,
Germany 2005) p. 359.

\bibitem{Dimopoulos} S. Dimopoulos {\em et al.}, 
\prl {\bf 98,} 111102 (2007); e-print arXiv:0802.4098 (2008).

\bibitem{Neutr} C. L\"ammerzahl, A. Macias, and H. M\"uller, Phys. Rev. A  {\bf 75,} 052104 (2007);
A. Arvanitaki {\em et al}, Phys. Rev. Lett. {\bf 100,} 120407
(2008).

\bibitem{Paris} H. M\"uller {\em et al.,}
Appl. Phys. B {\bf 84,} 633 (2006).


\bibitem{Hecker} J. Hecker-Denschlag {\em et al.}, J. Phys. B {\bf 35,} 3095 (2002).

\bibitem{CladeBOInterferometer} P. Clad\'e {\em et al.}, e-print
arXiv:0903.3511v1 (2009).

\bibitem{HauganKovachy} J. Hogan and T. Kovachy, private
communications.

\bibitem{beatfreq} For simplicity, the intensity as averaged over a period of the beat note
between the two frequencies is shown.


\bibitem{ellipfit} G. T. Foster, J. B. Fixler, J. M. McGuirk, and M. A.
Kasevich, Opt. Lett. {\bf 27,} 951 (2002).

\bibitem{Stockton} J. K. Stockton, X. Wu, and M. A. Kasevich, \pra
{\bf 76,} 033613 (2007).






\bibitem{Treutlein} P. Treutlein, K. Y. Chung, and S. Chu, Phys. Rev. A {\bf 63,} 051401(R)
(2001).







\bibitem{6Wlaser} S.-w. Chiow {\em et al.}, Optics
Express {\bf 17,} 5246 (2009).

\end{references}
\end{document}